\documentclass[aps,preprint]{revtex4}%
\usepackage{amsfonts}
\usepackage{amsmath}
\usepackage{amssymb}
\usepackage{graphicx}%
\begin{document}
\title{{\LARGE IS THE UNIVERSE A WHITE-HOLE?}}
\author{Marcelo Samuel Berman$^{1}$}
\affiliation{$^{1}$Instituto Albert Einstein\ - Av. Candido Hartmann, 575 - \ \# 17}
\affiliation{80730-440 - Curitiba - PR - Brazil}
\date{(Last Version) \ 31 March 2007.}

\begin{abstract}
Pathria(1972) has shown, for a pressureless closed Universe, that it is inside
a black (or white) hole. We show now, that the Universe with a cosmic pressure
obeying Einstein's field equations, \ can be inside a white-hole. In the
closed case, a positive cosmological constant does the job; for the flat and
open cases, the condition we find is not verified for the very early Universe,
but with the growth of the scale-factor, the condition will be certainly
fulfilled for a positive cosmological constant, after some time. We associate
the absolute temperature of the Universe, with the temperature of the
corresponding white-hole.

\end{abstract}
\maketitle

\begin{center}
{\LARGE IS THE UNIVERSE A WHITE-HOLE?}

\bigskip

Marcelo Samuel Berman
\end{center}

\bigskip

\bigskip{\Large I. Introduction}

\bigskip The contents of this paper, submitted first in 24 May, 2006, to a
journal, is based on a method published by Gomide and Berman(1987); we are not
sure that since 1987, the present calculation has not been published by
someone else. 

\ 

\bigskip Pathria(1972) published a paper showing that, for certain values of
the cosmological constant, a pressureless finite (positively curved) Universe
could be inside a white-hole, if it obeyed Robertson-Walker's metric (with
\ $k=1$):

\bigskip

$ds^{2}=dt^{2}-\frac{R^{2}(t)}{\left[  1+\left(  \frac{kr^{2}}{4}\right)
\right]  ^{2}}d\sigma^{2}$ \ \ \ \ \ \ \ \ \ \ \ \ \ \ , \ \ \ \ \ \ \ \ \ \ \ \ \ \ \ \ \ \ \ \ \ \ \ \ \ \ \ \ (1)

\bigskip

where,

\bigskip

$d\sigma^{2}=dx^{2}+dy^{2}+dz^{2}$ \ \ \ \ \ \ \ \ \ \ \ \ . \ \ \ \ \ \ \ \ \ \ \ \ \ \ \ \ \ \ \ \ \ \ \ \ \ \ \ \ \ \ \ \ \ \ \ \ \ (2)

\bigskip

If \ $R$\ \ stands for the radius of a "large sphere"\ \ of mass \ $M$\ \ ,
while \ \ $G$\ \ \ stands for Newton's gravitational constant, a white-hole
could be described by the condition that \ $R$\ \ is smaller than its
Schwarzschild's radius:

\bigskip

$R<2GM$\ \ \ \ \ \ \ \ \ \ \ . \ \ \ \ \ \ \ \ \ \ \ \ \ \ \ \ \ \ \ \ \ \ \ \ \ \ \ \ \ \ \ \ \ \ \ \ \ \ \ \ \ \ \ \ \ \ \ \ \ \ \ \ \ \ \ \ \ \ (3)

\bigskip

It is known, on experimental grounds, that the present Universe, obeys the
Machian relation by Brans and Dicke(1961),

\bigskip

$R\sim GM$\ \ \ \ \ \ \ \ \ \ . \ \ \ \ \ \ \ \ \ \ \ \ \ \ \ \ \ \ \ \ \ \ \ \ \ \ \ \ \ \ \ \ \ \ \ \ \ \ \ \ \ \ \ \ \ \ \ \ \ \ \ \ \ \ \ \ \ \ \ \ \ (4)

\bigskip

We, thus, can think that the Universe may be inside a white-hole, but this
should be analyzed by means of Einstein's field equations for the above metric,

\bigskip

$\kappa\rho=3H^{2}+\frac{3k}{R^{2}}+\Lambda$ \ \ \ \ \ \ \ \ \ \ \ \ \ \ , \ \ \ \ \ \ \ \ \ \ \ \ \ \ \ \ \ \ \ \ \ \ \ \ \ \ \ \ \ \ \ \ \ \ \ \ \ \ (5)

\bigskip

$\kappa p=-2\frac{\ddot{R}}{R}-H^{2}-\frac{k}{R^{2}}-\Lambda$
\ \ \ \ \ \ \ \ \ \ \ \ \ \ , \ \ \ \ \ \ \ \ \ \ \ \ \ \ \ \ \ \ \ \ \ \ \ \ \ \ \ \ \ (6)

\bigskip

where,

\bigskip

$H=\frac{\dot{R}}{R}$ \ \ \ \ \ \ \ \ \ \ \ \ \ \ \ ,

\bigskip

and,

\bigskip

$\kappa=8\pi G$ \ \ \ \ \ \ \ \ \ \ \ \ \ \ \ .

\bigskip

We study below, all three tricurvature cases ( $k=0,$ $\pm1$ \ ). We also
treat the absolute temperature of black holes and the Universe.

\bigskip

Berman(2006;2006a) has studied the same subject by means of Mach's Principle,
and the hypothesis of a zero-total-energy for the Universe.

\bigskip

{\LARGE \bigskip II. Closed Universe as a White-hole}

\bigskip Considering a positive cosmological constant ($\Lambda>0$), and
remembering that the volume \ $V$\ \ , for a closed Robertson-Walker's
Universe, is given by(Adler, Bazin and Schiffer, 1975):

\bigskip

$V=2\pi^{2}R^{3}$\ \ \ \ \ \ \ \ \ \ \ \ \ \ \ \ ,\ \ \ \ \ \ \ \ \ \ \ \ \ \ \ \ \ \ \ \ \ \ \ \ \ \ \ \ \ \ \ \ \ \ \ \ \ \ \ \ \ \ \ \ \ \ \ \ \ \ \ \ \ \ \ \ \ (7)

\bigskip

we find,

\bigskip

$\rho=\frac{M}{2\pi^{2}R^{3}}$ \ \ \ \ \ \ \ \ \ \ \ \ \ \ \ \ \ . \ \ \ \ \ \ \ \ \ \ \ \ \ \ \ \ \ \ \ \ \ \ \ \ \ \ \ \ \ \ \ \ \ \ \ \ \ \ \ \ \ \ \ \ \ \ \ \ \ \ \ \ \ \ \ \ (8)

\bigskip

\bigskip From equation(5) we find:

\bigskip

$\kappa\rho=\frac{3\left[  1+\dot{R}^{2}\right]  }{R^{2}}+\Lambda=\frac{\kappa
M}{2\pi^{2}R^{3}}=\frac{2r}{\pi R^{3}}$
\ \ \ \ \ \ \ \ \ \ \ \ \ \ \ \ \ \ \ , \ \ \ \ \ \ \ \ \ \ \ \ \ \ \ \ \ \ \ \ \ \ \ \ \ \ (9)

\bigskip

where \ \ $r$\ \ \ stands for \ Schwarzschild's radius, i.e.,

\bigskip

$r=2GM=R\left\{  \frac{\pi}{2}\left[  3\dot{R}^{2}+3+\Lambda R^{2}\right]
\right\}  >R$\ \ \ \ \ \ \ \ . \ \ \ \ \ \ \ \ \ \ \ \ \ \ \ \ \ \ \ \ (10)

\bigskip

Because \ $\dot{R}^{2}>0$\ \ \ because \ $R(t)$\ \ is a real number, we find,

\bigskip

$\Lambda>\left[  \frac{2}{\pi}-3\right]  R^{-2}>-3R^{-2}$%
\ \ \ \ \ \ \ \ \ \ \ \ \ \ . \ \ \ \ \ \ \ \ \ \ \ \ \ \ \ \ \ \ \ \ \ \ \ \ \ \ \ \ \ \ \ \ \ \ \ \ \ \ (11)

\bigskip

\bigskip As the scale-factor is essentially positive, condition (11) will be
certainly fulfilled by the condition, \ $\Lambda>0$\ \ .\ 

We have shown that a closed Universe, while cosmic pressure \ $p$\ \ follows
Einstein's field equation (6) normally, shall be inside a white-hole, if
\ \ $\Lambda>0$\ .

{\Large III. Flat Universe as a White-hole}

\bigskip

We shall now apply Einstein's field equations for \ $k=0$\ \ . With the prescription:

\bigskip

$V=\frac{4}{3}\pi R^{3}$\ \ \ \ \ \ \ \ \ \ \ \ \ , \ \ \ \ \ \ \ \ \ \ \ \ \ \ \ \ \ \ \ \ \ \ \ \ \ \ \ \ \ \ \ \ \ \ \ \ \ \ \ \ \ \ \ \ \ \ \ \ \ \ \ \ \ \ \ \ \ \ (12)

\bigskip

so that,

\bigskip

$\rho=\frac{3M}{4\pi R^{3}}$ \ \ \ \ \ \ \ \ \ \ \ \ \ \ , \ \ \ \ \ \ \ \ \ \ \ \ \ \ \ \ \ \ \ \ \ \ \ \ \ \ \ \ \ \ \ \ \ \ \ \ \ \ \ \ \ \ \ \ \ \ \ \ \ \ \ \ \ \ \ \ \ \ \ (13)

\bigskip

we would find,

\bigskip

$r=R^{3}\left[  H^{2}+\frac{\Lambda}{3}\right]  =R\left[  \dot{R}^{2}%
+\frac{\Lambda}{3}R^{2}\right]  $ \ \ \ \ \ \ \ \ \ \ \ \ \ . \ \ \ \ \ \ \ \ \ \ \ \ \ \ \ \ \ \ \ \ \ \ (14)

\bigskip

We find that a sufficient condition for \ \ $r>R$\ \ is,

\bigskip

$\Lambda>\frac{3}{R^{2}}$\ \ \ \ \ \ \ \ \ \ . \ \ \ \ \ \ \ \ \ \ \ \ \ \ \ \ \ \ \ \ \ \ \ \ \ \ \ \ \ \ \ \ \ \ \ \ \ \ \ \ \ \ \ \ \ \ \ \ \ \ \ \ \ \ \ \ \ \ \ \ \ \ \ \ \ \ \ \ \ (15)

\bigskip

We remark that this condition is equivalent to "closing" the Universe, in what
respects the density field equation (but not in terms of the pressure
equation). Even if this condition may not prevail for the very early Universe,
nevertheless, with increasing values of the scale-factor, eventually the
condition (15) shall be fulfilled.

\bigskip

The limiting scale-factor, is of course,

\bigskip

$R>R_{0}=\sqrt{\frac{3}{\Lambda}}$ \ \ \ \ \ \ \ \ \ \ \ \ \ \ \ . \ \ \ \ \ \ \ \ \ \ \ \ \ \ \ \ \ \ \ \ \ \ \ \ \ \ \ \ \ \ \ \ \ \ \ \ \ \ \ \ \ \ \ \ \ \ \ \ \ \ (16)

\bigskip{\Large IV. Absolute temperature in Cosmology and White-holes}

\bigskip Hawking(2001), cites the formula he obtained many years ago, for the
absolute temperature \ $T$\ \ associated with a non rotating and non charged
black-hole; though in our paper we are dealing with Classical Physics, and
Hawking's formula is based on Quantum theory, it will be seen that we can make
a bridge with Classical Cosmology:

\bigskip

$T=\frac{\bar{h}\text{ }c^{3}}{8\pi kR_{S}}$ \ \ \ \ \ \ \ \ \ \ \ \ \ \ \ , \ \ \ \ \ \ \ \ \ \ \ \ \ \ \ \ \ \ \ \ \ \ \ \ \ \ \ \ \ \ \ \ \ \ \ \ \ \ \ \ \ \ \ \ \ \ \ (17)

\bigskip

where, $k$\ \ , and \ $h$\ \ stand respectively for Boltzman and Planck
constants, and \ \ $R_{S}$\ \ \ stands for Schwarzschild's radius, in \ such a
way that, \ if we associate the scale-factor with \ \ $R_{S}$\ \ \ \ , then,
from (17),

\bigskip

$T\cong$\ $\frac{hc^{3}}{16\pi^{2}k}R^{-1}$ \ \ \ \ \ \ \ \ \ \ \ \ \ \ \ . \ \ \ \ \ \ \ \ \ \ \ \ \ \ \ \ \ \ \ \ \ \ \ \ \ \ \ \ \ \ \ \ \ \ \ \ \ \ \ \ \ \ \ \ \ \ \ (18)

\bigskip

This is exactly what we \ find in standard Cosmology, \ between \ the scale
factor \ $R(t)$\ and \ $T$\ \ .\ This is a most striking analogy that points
out to a possible white-hole Universe. However, Berman(2006 b; 2007), found
that, either for the Machian Universe, or for a 4-D black hole, the entropy is
given by the same formula, i.e.,

$S\propto R^{3/2}$\ \ \ \ \ ,

\bigskip

provided that, the scale factor and Schwarzschild's radius, obey the condition,

\bigskip

\ $R\propto T^{-2}$ \ \ \ \ .

\bigskip

\bigskip The above condition, for the temperature, is a Machian Universe
property (Berman, 2006 b, 2007).

\ \ \ 

\bigskip{\Large V. Open Universe as a white-hole}

\bigskip

We shall leave, once more, the cosmic pressure to obey freely Einstein's field
equations. From the energy density \ $\rho$\ \ -- equation (5), with
\ $k=-1$\ \ , we have, then,

\bigskip

$r=R\left[  \frac{\alpha}{4\pi}\left(  3R^{2}H^{2}+\Lambda R^{2}-3\right)
\right]  $ \ \ \ \ \ \ \ \ \ \ \ \ \ \ \ \ \ , \ \ \ \ \ \ \ \ \ \ \ \ \ \ \ \ \ \ (19)

\bigskip

where we suppose, quite generally, that the tridimensional volume is given by:

\bigskip

$V=\alpha R^{3}$ \ \ \ \ \ \ \ \ \ \ \ \ \ \ \ \ \ ( \ $\alpha=$\ positive
constant \ ) \ \ \ . \ \ \ \ \ \ \ \ \ (20)

\bigskip

The condition for the open Universe to be or become a white-hole, is given by
\ \ $r>R$\ \ , so that,

\bigskip

$\Lambda>\left[  m-3\dot{R}^{2}\right]  R^{-2}$
\ \ \ \ \ \ \ \ \ \ \ \ \ \ \ \ , \ \ \ \ \ \ \ \ \ \ \ \ \ \ \ \ \ \ \ \ \ \ \ \ \ \ \ \ \ \ \ \ \ \ \ \ \ \ \ (21)

\bigskip

where,

\bigskip

$m=3+\frac{4\pi}{\alpha}>3$ \ \ \ \ \ \ \ \ \ \ \ \ \ \ \ \ \ \ \ \ \ . \ \ \ \ \ \ \ \ \ \ \ \ \ \ \ \ \ \ \ \ \ \ \ \ \ \ \ \ \ \ \ \ \ \ \ \ \ \ (22)

\bigskip

We conclude that the condition (21) can be staten as,

\bigskip

$\Lambda>\left[  3(1-\dot{R}^{2})\right]  R^{-2}$
\ \ \ \ \ \ \ \ \ \ \ \ \ \ \ \ \ . \ \ \ \ \ \ \ \ \ \ \ \ \ \ \ \ \ \ \ \ \ \ \ \ \ \ \ \ \ \ \ \ \ \ \ \ \ (23)

\bigskip

At least, we can impose the following sufficient condition:

\bigskip

$\Lambda>3R^{-2}>0$ \ \ \ \ \ \ \ \ \ \ \ \ \ \ \ \ \ , \ \ \ \ \ \ \ \ \ \ \ \ \ \ \ \ \ \ \ \ \ \ \ \ \ \ \ \ \ \ \ \ \ \ \ \ \ \ \ \ \ \ \ \ \ \ (24)

\bigskip

because \ $\dot{R}^{2}>0$\ \ \ .

\bigskip

By the same token as in last paragraph of Section III above, for the expanding
Universe, when the scale factor becomes larger than \ \ $\sqrt{\frac
{3}{\Lambda}}$\ \ \ we shall have a white-hole-Universe.

\bigskip

We remark that condition (24) is identical with (16), which refers to a flat
Robertson-Walker's metric.

\bigskip

{\Large VI. Conclusions}

\bigskip

We have shown, in a different context than in Pathria's paper (where \ $p=0$ ,
and \ \ $\Lambda$\ \ obeys certain conditions), that the closed
Robertson-Walker's Universe, with any value of \ $p$\ constrained to obey
Einstein's field equations may be thought as being a white-hole. Brans-Dicke
relation to the problem is not conclusive, because it represents only a
Machian condition for the Universe. In a similar way, flat or open Universes,
in the expanding phase, may become white-holes after $R$\ \ becoming
\ \ larger than \ $\sqrt{\frac{3}{\Lambda}}$\ . Finally, we motivate the
analogy between the Universe and a white-hole, by means of their absolute
temperatures. The Machian condition for the Universe, was taken, \ partially,
as implying that the absolute temperature ran like \ $\left(  \sqrt{R}\right)
^{-1}$\ \ . (Berman, 2006 b, 2007)\ . \ The standard condition, however, is
that \ $T\propto R^{-1}$\ \ , like we considered above.

\ \ \ 

{\Large Acknowledgements}

\bigskip

The author gratefully thanks his intellectual mentors, Fernando de Mello
Gomide and M. M. Som, and is also grateful for the encouragement by Geni,
Albert, and Paula, and for the useful comments by Dimi Chakalov.

\bigskip

{\Large References}

\bigskip

Adler, R.; Bazin, M,; Schiffer, M. (1975) - \textit{Introduction to General
Relativity}\ - 2nd. edtn., McGraw-Hill, N.Y.

\bigskip Berman,M.S. (2006) - \textit{Energy of Black-Holes and Hawking's
Universe \ }in \textit{Trends in Black-Hole Research, }Chapter 5\textit{.}
Edited by Paul Kreitler, Nova Science, New York.

Berman,M.S. (2006 a) - \textit{Energy, Brief History of Black-Holes, and
Hawking's Universe }in \textit{New Developments in Black-Hole Research},
Chapter 5\textit{.} Edited by Paul Kreitler, Nova Science, New York.

Berman,M.S. (2006 b) - On the Machian Properties of the Universe, submitted to
publication, Los Alamos Archives, physics/0610003 v3.

Berman,M.S. (2007) - \textit{Introduction to General Relativity, and the
Cosmological Constant Problem}, Nova Science, New York. [Just published!!].

Brans, C.; Dicke, R.H. (1961) - Physical Review, \textbf{124}, 925.

\bigskip Gomide, F. M.; Berman, M.S. (1987) - \textit{Introduction to
Relativistic Cosmology} (in Portuguese), McGraw-Hill, S.P.

\noindent Hawking, S. (2001) - \textit{The Universe in a Nutshell}, Bantam,
New York.

Pathria, R.K. (1972) - Nature \textbf{240}, 298.

\end{document}